\documentclass{PoS}
\usepackage{array}

\usepackage{slashed}

\newcommand{\be}{\begin{equation}}
\newcommand{\ee}{\end{equation}}
\newcommand{\bea}{\begin{eqnarray}}
\newcommand{\eea}{\end{eqnarray}}

\title{Gluon transverse momentum dependent correlators in polarized high energy processes}

\ShortTitle{Gluon TMD correlators}

\author{Daniel Boer\\
        Van Swinderen Institute, University of Groningen,\\ Nijenborgh 4, NL-9747 AG Groningen, Netherlands\\
        E-mail: \email{d.boer@rug.nl}}

\author{Sabrina Cotogno\\
        Nikhef Theory Group and Department of Physics and Astronomy, VU University Amsterdam\\
        De Boelelaan 1081, NL-1081 HV Amsterdam, the Netherlands\\
        E-mail: \email{scotogno@nikhef.nl}}

\author{Tom van Daal\\
        Nikhef Theory Group and Department of Physics and Astronomy, VU University Amsterdam\\
        De Boelelaan 1081, NL-1081 HV Amsterdam, the Netherlands\\
        E-mail: \email{tvdaal@nikhef.nl}}

\author{\speaker{PIET J. MULDERS}\\
        Nikhef Theory Group and Department of Physics and Astronomy, VU University Amsterdam\\
        De Boelelaan 1081, NL-1081 HV Amsterdam, the Netherlands\\
        E-mail: \email{p.j.g.mulders@vu.nl}}

\author{Andrea Signori\\
        Nikhef Theory Group and Department of Physics and Astronomy, VU University Amsterdam\\
        De Boelelaan 1081, NL-1081 HV Amsterdam, the Netherlands\\
        E-mail: \email{asignori@nikhef.nl}}

\author{Yajin Zhou\\
        School of Physics \& Key Laboratory of Particle Physics and Particle Irradiation (MOE),\\ Shandong University, Jinan, Shandong 250100, China\\
        E-mail: \email{zhouyj@sdu.edu.cn}}

\vspace{1cm}

\abstract{We investigate the gluon transverse momentum dependent correlators as Fourier transform of matrix elements of nonlocal operator combinations. At the operator level these correlators include both field strength operators and gauge links bridging the nonlocality. In contrast to the collinear PDFs, the gauge links are no longer unique for transverse momentum dependent PDFs (TMDs) and also Wilson loops lead to nontrivial effects. We look at gluon TMDs for unpolarized, vector and tensor polarized targets. In particular a single Wilson loop operators become important when one considers the small-x limit of gluon TMDs.
}

\FullConference{XXIV International Workshop on Deep-Inelastic Scattering and Related Subjects\\
                 11 - 15 April 2016,
                 DESY, Hamburg}

\begin{document}

\section{Introduction}
Parton distribution functions (PDFs) establish the connection between hadrons in initial state and the hard process. As such they replace in the basic description of the cross section the polarization sums for quarks and gluons by {\em correlators},
\bea 
&&u_i(k)\overline u_j(k) \ \propto\  \slashed{k}_{ij} \ \Longrightarrow \ \Phi_{ij}(k;P,S),\\
&&\epsilon^{\alpha}(k)\epsilon^{\beta\ast}(k) \ \propto\ -g_T^{\alpha\beta} \ \Longrightarrow \ \Gamma^{\alpha\beta}(k;P)
\eea
for a quark with momentum $k$ in a hadron with momentum $P$. In high energy processes the hadron momenta defining light-like directions and we use the light-like vector $n$ with $P{\cdot}n = 1$. The correlators involve quark and gluon fields. The important combination of fields, referred to as {\em leading twist}, are those minimizing the canonical dimension. We use a Sudakov expansion of the parton momentum, $k = x\,P + k_T + \ldots$, where $\ldots$ is along $n$ and is the irrelevant momentum component that can be integrated over leaving correlators $\Phi_{ij}(x,k_T)$ and $\Gamma^{\alpha\beta}(x,k_T)$ for quarks and gluons, respectively. Including transverse momenta of the partons they are parametrized in structures with specific Dirac ($ij$) and Lorentz structure ($\alpha\beta$) and transverse momentum dependent (TMD) PDFs $f_{\ldots}(x,k_T^2)$, in short referred to as TMDs.   

The high energy kinematics is an essential ingredient in this. In the center of mass of the partonic scattering process, the hadronic momenta are in essence light-like and the hadronic masses becomes irrelevant. In the hadron rest frame, a given hadron is struck at one particular light-front time. In this situation the light-cone fractions $x$ of the parton momentum can be identified with scaling variables, such as the Bjorken scaling variable $x_B = -q^2/2P{\cdot}q$ in deep inelastic scattering. The transverse components of the parton with respect to the hadron can be accessed in processes where one observes a non-collinearity in at least three hard momenta, that in the absence of intrinsic transverse momentum in the hadrons are expected to be collinear, of course given a particular partonic subprocess.  

This inclusion of transverse momentum and TMD factorization all works straightforward at tree-level with factorization theorems being studied. Effects of {\em intrinsic} transverse momenta of partons are best visible in (partially) polarised processes. In that case one has polarization vectors or tensors for hadrons (parametrizing the spin density matrix) or measurable polarization vectors depending on final state distributions of decay products. The initial state spin (symmetric and traceless) vectors or tensors for hadrons are in analogy to the momentum expanded as
\bea
S^\mu &=& S_L \frac{P^\mu}{M} + S_T^\mu - MS_L \,n^\mu, 
\nonumber \\
S^{\mu\nu} &=& \frac{1}{2} \left[ \frac{2}{3} S_{LL} \,g_T^{\mu\nu} + \frac{4}{3} S_{LL} \frac{P^\mu P^\nu}{M^2} + \frac{S_{LT}^{\{\mu}P^{\nu\}}}{M} + S_{TT}^{\mu\nu} 
- \,\frac{4}{3} S_{LL} P^{\{\mu}n^{\nu\}} - M S_{LT}^{\{\mu}n^{\nu\}} + \frac{4}{3} M^2 S_{LL} \,n^\mu n^\nu \vphantom{\frac{P^\mu P^\nu}{M^2}} \right] ,
\nonumber
\eea
ensuring the relations $P {\cdot} S = P_\mu S^{\mu\nu} = 0$.

\section{TMD correlators and distribution functions}

The quark and gluon TMD correlators in terms of matrix elements of quark fields~\cite{Collins:1981uw,Collins:1981uk,Collins:2011zzd} including the Wilson lines $U$ needed for color gauge invariance of the TMD case are given by
\begin{eqnarray}
&&
\Phi_{ij}^{[U]}(x,p_T;n)
=\int \frac{d\,\xi{\cdot}P\,d^{2}\xi_T}{(2\pi)^{3}}
\,e^{ip\cdot \xi} \langle P{,}S\vert\overline{\psi}_{j}(0)
\,U_{[0,\xi]}\psi_{i}(\xi)\vert P{,}S\rangle\,\big|_{LF},
\label{e:operator}
\\&&
\Gamma^{[U,U^\prime]\,\mu\nu}(x,p_T;n) ={\int}\frac{d\,\xi{\cdot}P\,d^2\xi_T}{(2\pi)^3}\ e^{ip\cdot\xi}
\,\langle P{,}S\vert\,F^{n\mu}(0)\,U_{[0,\xi]}^{\phantom{\prime}}\,F^{n\nu}(\xi)\,U_{[\xi,0]}^\prime\,\vert P{,}S\rangle\big|_{LF}
\end{eqnarray}
(color summation or color tracing implicit). The non-locality in the integration is limited to the lightfront, $\xi{\cdot}n = 0$, indicated with LF. The gauge links $U_{[0,\xi]}^{\phantom{\prime}}$ are path ordered exponentials (two different ones for gluons) needed to make the correlator gauge invariant~\cite{Belitsky:2002sm,Boer:2003cm,Bomhof:2006dp}. For the quark correlator the gauge link bridges the non-locality, which in the case of TMDs involves also transverse separation. The simplest ones are the future- and past-pointing staple links $U_{[0,\xi]}^{[\pm]}$ (or just $[\pm]$) that just connect the points $0$ and $\xi$ via lightcone plus or minus infinity, $U_{[0,\xi]}^{[\pm]} = U_{[0,\pm\infty]}^{[n]} U_{[0_{\scriptscriptstyle T},\xi_{\scriptscriptstyle T}]}^{{\scriptscriptstyle T}}U_{[\pm \infty,\xi]}^{[n]}$. We use these as our basic building blocks. For gluons TMDs the most general structure involves two gauge links (triplet representation), denoted as $[U,U^\prime]$, connecting the positions $0$ and $\xi$ in different ways. The simplest combinations allowed for $[U,U^\prime]$ are $[+,+]$, $[-,-]$, $[+,-]$ and $[-,+]$. More complicated possibilities, e.g. with additional (traced) Wilson loops of the form $U^{[\square]}=U_{[0,\xi]}^{[+]}U_{[\xi,0]}^{[-]}$ = $U_{[0,\xi]}^{[+]}U_{[0,\xi]}^{[-]\dagger}$ or its conjugate are allowed as well. A list with all type of contributions can be found in Ref.~\cite{Buffing:2012sz,Buffing:2013kca}. If $U = U^\prime$ one can also use a single gauge link in the octet representation.

We note that the combination of two different gauge lines as appearing in the gluon correlator even without explicit gluon fields yields an interesting nonvanishing matrix element appearing in the correlator
\begin{equation}
\Gamma_0^{[+,-]}(x,p_T;n) \equiv  \int \frac{d\xi{\cdot} P\,d^2\xi_T}{(2\pi)^3} \,e^{ip{\cdot}\xi} \langle P,S\vert U_{[0,\xi]}^{[+]} U_{[\xi,0]}^{[-]} \vert P,S \rangle\big|_{LF}  = \delta(x)\,\Gamma_0^{[+,-]}(p_T;n).
\label{e:gamma0_up}
\end{equation}

None of the above correlators can be calculated from first principles. They are parametrized in terms of TMD PDFs, which at the level of leading twist contributions for unpolarized hadrons is given by
\begin{eqnarray}
\Phi^{[U]}(x,p_{T};n)&=&\bigg\{
f^{[U]}_{1}(x,p_T^2)  + i\,h_1^{\perp [U]}(x,p_T^2)\frac{\slashed{p}_T}{M}
\bigg\}\frac{\slashed{P}}{2} ,
\label{e:QuarkCorr}
\\
\Gamma^{\mu\nu [U,U^\prime]}(x{,}p_T) &=& 
-g_T^{\mu\nu}\,f_1^{g [U]}(x{,}p_{\scriptscriptstyle T}^2)
+\bigg(\frac{p_T^\mu p_T^\nu}{M^2}\,{-}\,g_T^{\mu\nu}\frac{p_{\scriptscriptstyle T}^2}{2M^2}\bigg)\;h_1^{\perp g [U]}(x{,}p_{\scriptscriptstyle T}^2).
\label{e:GluonCorr}
\end{eqnarray}
For quarks and gluons one can find results in Refs~\cite{Mulders:1995dh,Bacchetta:2006tn,Mulders:2000sh,Meissner:2007rx}, including tensor polarized targets for quarks. 
The gauge link dependence in these parametrizations is contained in the TMDs. Note that for quarks $h_1^\perp$ is T-odd, while for gluons both functions are T-even. At this meeting we report on the parametrization including tensor polarization in detail outlined in Ref.~\cite{Boer:2016xqr}, including also the parametrization of the Wilson loop correlator that only depends on $p_T$ and for unpolarized hadrons just is proportional to a scalar function.
\begin{equation}
\Gamma_0^{[+,-]}(p_T) \propto e(p_T^2).
\end{equation} 

Even if any gauge link defines a gauge invariant correlator, the relevant gauge links to be used in a given process just follow from a correct resummation of all diagrams including the exchange of any number of $A^n$ gluons between the hadronic parts and the hard part, i.e.\ gluons with their polarization along the hadronic momentum. They nicely sum to the path-ordered exponential. For quark distributions in semi-inclusive deep inelastic scattering they resum into a future-pointing gauge link, in the Drell-Yan process they resum into a past-pointing gauge link, which is directly linked to the color flow in these processes. Color flow arguments suggest that the Wilson loop correlator may be important in diffractive processes.

\section{Operator analysis}

In the situation of collinear PDFs (integrated over transverse momenta), the non-locality is restricted to the lightcone, $\xi{\cdot}n = \xi_T = 0$ (LC) and the staple links reduce to straight-line Wilson lines. The correlators then involve the non-local operator combinations $\overline\psi(0)U^{[n]}_{[0,\xi]}\psi(\xi)\vert_{LC}$ or $F^{n\mu}(0)U^{[n]}_{[0,\xi]}F^{n\nu}(\xi)U^{[n]}_{[\xi,0]}\vert_{LC}$, expanded in terms of leading twist operators $\overline\psi(0)D^n\ldots D^n\psi(0)$ and ${\rm Tr}[F^{n\mu}D^n\ldots D^n F^{n\nu}(0)D^n\ldots D^n]$ operators. For transverse momentum dependent correlators the dependence on $\xi_T$ gives in the parametrization distribution functions multiplied with Dirac or Lorentz structures involving $p_T$. It is useful to look at the $p_T$ dependence in terms of symmetric and traceless tensors (such as for instance in the parametrization of the gluon correlator). The rank of the tensor defines the rank of the distribution function. Rank zero functions are the collinear PDFs. 

Considering distribution functions of definite rank is also useful because the functions have the same rank in impact parameter space, which is important for the study of evolution. In principle a factor $p_T$ in the parametrization can be rewritten as a derivative working on $\xi_T$. For distribution functions such differentiation gives rise to two types of operators in the correlator, for quarks being of the form 
\bea
&&
\widetilde\Phi_{\hat O,ij}^{[U]}(x,p_{T})
=\int \frac{d\,\xi{\cdot}P\,d^{2}\xi_{T}}{(2\pi)^{3}}
\,e^{ip\cdot \xi} \langle P{,}S\vert\overline{\psi}_{j}(0)
\,U_{[0,\xi]}\hat O(\xi)\psi_{i}(\xi)\vert P{,}S\rangle\,\Big|_{LF},
\label{e:twistoperatorquark}
\eea
where the $\hat O(\xi)$ operators are combinations of $i\partial_T(\xi) = iD_T^\alpha(\xi) - A_T^\alpha(\xi)$ and $G^\alpha(\xi)$, defined in a color gauge invariant way (thus including gauge links),
\bea
&&A_{T}^{\alpha}(\xi)=\frac{1}{2}\int_{-\infty}^{\infty}
d\eta{\cdot}P\ \epsilon(\xi{\cdot}P-\eta{\cdot}P)
\,U_{[\xi,\eta]}^{[n]} F^{n\alpha}(\eta)U_{[\eta,\xi]}^{[n]}, 
\label{e:defA} 
\\
&&G^{\alpha}(\xi)=\frac{1}{2}\int_{-\infty}^{\infty}
d\eta{\cdot}P\ U_{[\xi,\eta]}^{[n]}F^{n\alpha}(\eta)
U_{[\eta,\xi]}^{[n]},
\label{e:defG}
\eea
with $\epsilon (\zeta)$ being the sign function. Note that $G^{\alpha}(\xi)$ = $G^{\alpha}(\xi_T)$ does not depend on $\xi{\cdot}P$, implying in momentum space $p\cdot n = p^+ = 0$, hence the name gluonic pole matrix elements~\cite{Efremov:1981sh,Efremov:1984ip,Qiu:1991pp,Qiu:1991wg,Qiu:1998ia,Kanazawa:2000hz}. The operator in Eq.~\ref{e:defG} is actually time-reversal odd, giving rise to leading terms in the correlators that are important for single spin asymmetries. The factors multiplying the gluonic pole correlators depend on the gauge links in the correlators, which as already mentioned only depend on the hard process in which the correlators are needed to connect to the hadrons involved. Most well-known are the single gluonic pole factors for staple links $C_G^{[\pm]} = \pm 1$ giving rise to the sign flip of T-odd distribution functions going from semi-inclusive deep inelastic scattering to Drell-Yan linked to the different color flow in the respective underlying hard processes.

Depending on the rank of the functions, more gluonic pole operators may enter~\cite{Boer:2015kxa}. For two gluonic poles this lead to an interesting link between the Wilson loop correlator and the gluon correlator at $x = 0$. The Wilson loop correlator only depends on $t$ in the small-x region where $k^2 \approx k_T^2$ and one finds for $x\approx 0$ 
\begin{equation}
\Gamma^{[+,-]\,\alpha\beta}(x,p_T) \propto \frac{p_T^\alpha p_T^\beta}{M^2} \,\Gamma_0^{[+,-]}(t).
\label{limit}
\end{equation}
The above results agree with the result in ref.~\cite{Dominguez:2011wm} where in the small-$x$ limit $f_1^{[+,-]}(x,p_T^2)$ becomes proportional to the dipole cross section. In Ref.~\cite{Boer:2015pni} that connection was already made on the correlator level for the case of a transversely polarized nucleon. Results for tensor polarization are included in Ref.~\cite{Boer:2016xqr}. We note that in general combination gluonic poles operators $G^\alpha$ with the field strengths $F^{n\alpha}$ or the quark fields, can give rise to multiple color neutral combinations which has to be taken into account.

\section{Conclusions}

TMDs enrich the partonic structure of hadrons as compared to collinear PDFs. At the technical level, there are a number of complications such as the appropriate process-dependent gauge links, the matching of small and large $p_T$ and the more complex evolution that need to be addressed. The study of the operator structure of TMDs with definite rank is important and instructive to study the role of TMDs for polarized hadrons and for establishing links to small-x physics. 

\acknowledgments
This research is part of the research program of the ``Stichting voor Fundamenteel Onderzoek der Materie (FOM)'', which is financially supported by the ``Nederlandse Organisatie voor Wetenschappelijk Onderzoek (NWO)'' and the EU "Ideas" programme QWORK (contract 320389).


\end{document}